\documentclass[twocolumn]{aastex701}

\usepackage{multirow}
\usepackage{amsmath}

\usepackage{rotating}
\usepackage{afterpage}

\begin{document}

\title{\textit{Exploring Galactic open clusters with Gaia. II} Dynamical Evolution and Stellar Population Properties in Fifteen Nearby Open Clusters}

\author[orcid=0009-0004-5488-4945]{Jeison Alfonso}
\affiliation{Universidad Cat\'{o}lica de la Sant\'{i}sima Concepci\'{o}n,
          Departamento de Ingenier\'{i}a El\'{e}ctrica,
          Alonso de Ribera 2850,
          Concepci\'{o}n, Chile}
\affiliation{Universidad de los Andes,
          Departamento de F\'{i}sica,
          Cra. 1 No. 18A-10, Bloque Ip, A.A. 4976
          Bogot\'{a}, Colombia}
\email[show]{jalfonso@ucsc.cl}

\author[orcid=0000-0001-8351-0628]{Alejandro Garc\'{i}a-Varela} 
\affiliation{Universidad de los Andes,
          Departamento de F\'{i}sica,
          Cra. 1 No. 18A-10, Bloque Ip, A.A. 4976
          Bogot\'{a}, Colombia}
\email{josegarc@uniandes.edu.co}

\author[orcid=0000-0001-5598-8720]{Katherine Vieira}
\affiliation{Instituto de Astronom\'{i}a y Ciencias Planetarias, 
          Universidad de Atacama, Copayapu 485, 
          Copiap\'{o} 1531772, Chile}
\email{katherine.vieira@uda.cl}

%\collaboration{all}{The Terra Mater collaboration}

\begin{abstract}
Stellar mass governs stellar evolution, and the distribution of stellar masses plays a central role in the dynamical evolution of stellar clusters. 
Using high-precision astrometry and photometry from Gaia DR3, we investigate mass segregation and the present-day mass function (PDMF) in fifteen nearby open clusters. 
Single and binary stars are identified from the color–magnitude diagram, and stellar masses for single stars are derived from a cubic spline relation with $G$-band magnitude, while binary component masses are estimated via a simulation-based inference method. 
Based on these mass estimates, mass segregation is quantified using the minimum spanning tree technique, and the PDMF is characterized through power-law fitting. 
We detect significant mass segregation in all clusters. Systems lacking very massive stars exhibit weaker segregation affecting a larger fraction of the population, whereas clusters hosting a small number of very massive stars show strong central concentration of these objects. 
The PDMF follows a power law with a slope of $(2.01\pm0.25)$, consistent with the canonical Kroupa initial mass function. 
The strength of mass segregation correlates with the mass of the most segregated stars.
Strong segregation observed in very young clusters supports a primordial origin, while older clusters display signatures of dynamical mass segregation at lower masses and evidence of binary disruption. 
A low-mass break in the PDMF observed in most clusters, if not due to incompleteness, may reflect early gas expulsion in initially
mass-segregated systems.

\end{abstract}



\section{Introduction} 

The Initial Mass Function (IMF), which characterizes the initial distribution of stellar masses resulting from a star formation event in the Milky Way \citep{salpeter1955}, stands as one of the fundamental pillars in the understanding of stellar populations and galactic evolution, serving as a crucial input parameter for models of stellar evolution, chemical enrichment, and galaxy formation. 
This foundational concept has evolved significantly since its introduction. 
\citet{millerscalo1979} provided one of the first analyses suggesting that the IMF might not follow a single power law across all masses. 
The theoretical framework supporting this suggestion was further developed through the multi-segment power-law proposed by \citet{kroupa2001} and the log-normal distribution suggested by \citet{chabrier2003}. 
The canonical Kroupa IMF is represented by two power laws: one with a slope of $\alpha = 1.3$ for stars with masses between $0.08$ and $0.5 M_{\odot}$, and another with a slope of $\alpha = 2.3$ for massive stars, up to the maximum stellar mass that can be formed in an embedded cluster \citep{Kroupa2024}. 
In contrast, the \citet{chabrier2003} IMF describes the observed mass distribution for stars with masses below $1 M_{\odot}$.

\citet{Kirkpatrick2024}, using Gaia DR3, analyzed detections that represent approximately 75$\%$ of the volume-complete census within 20 pc. 
Objects in this region may be missed by Gaia if they are too bright, too faint, or are companions to Gaia-detected stars with insufficient astrometric characterization. 
These authors have found that a quadripartite power-law model provides a good fit to the data. For the mass range from high to low, they report IMF exponents of $\alpha=$ 2.3, 1.3, 0.25, and 0.6, within the following mass ranges $(8, 0.55)$, $(0.55, 0.22)$, $(0.22, 0.05)$ and $(0.05, 0.01)$ $M_{\odot}$, respectively.

On the other hand, \citet{hennebelle2024} provided a review of the physical processes that determine the stellar IMF, highlighting how the interplay between gravity and turbulence is responsible for establishing the high-mass power-law slope. 
Furthermore, the formation of the first hydrostatic core may drive the peak of the IMF, primarily influenced by dust opacity and the physics of molecular hydrogen. 
This process is only weakly affected by factors such as radiation, magnetic fields, turbulence, and metallicity, among other environmental conditions.

The low-mass slope of IMF distribution depends in some measure on the number of binary (and multiple) stars present in the cluster. 
The IMF counts how many stars are formed per bin of mass, regardless of whether they are single or multiple. 
Therefore, the masses of the secondary (and higher order) companions in non-single stars may contribute significantly at the low mass level. 
When the absolute value of a binary system’s binding energy exceeds the average kinetic energy of encounters with third stars, the system becomes resilient to disruption through such interactions \citep{binney2008}. 
These systems, referred to as hard binaries, dynamically behave as single entities with an effective mass equal to the sum of their components, that is $m_1 + m_2$. 
Then, as time passes, stellar encounters will disrupt the binaries, being more effective with the so-called wide binaries \citep[e.g.][]{deacon2020}, while the harder ones survive, acting like a single more massive star. 
In some cases, these stellar encounters expel one of the binary components from the cluster, a dynamical ejection considered to be the source of field young massive stars \citep[e.g.][]{oh2015}.

Mass segregation, the tendency of more massive stars to concentrate toward the cluster center, is a key dynamical process that shapes the evolution of open clusters. 
It influences both the observed stellar mass function and the spatial distribution of stars. 
\citet{mcmillan2007} proposed that this phenomenon may be primordial, inherited from the hierarchical assembly of initially mass-segregated clumps that merge after undergoing two-body relaxation. 
In contrast, \citet{Haghi2015} demonstrated via $N$-body simulations that gas expulsion from initially mass-segregated clusters leads to a significantly shallower slope in the low-mass end of the stellar mass function.
Stellar multiplicity also plays a fundamental role in early cluster dynamics. 
\citet{Kroupa1995a} suggested that all stars form in binary or higher-order multiple systems. 
To explore the impact of binaries on primordial mass segregation, \citet{Pavlik2019} conducted $N$-body simulations and found that fully mass-segregated clusters, even when including binaries, reproduce the observed properties of the Orion Nebula Cluster more accurately than non-segregated ones.

Mechanisms such as tidal captures and the encounters of three individual stars have been proposed to explain the formation of binary systems, however these cannot explain the observed fraction of binary systems in open clusters (see \cite{Dabrin2022} and references therein).
Using ALMA data, \citet{Plunkett2018} found evidence of primordial mass segregation among low-mass protostellar sources in the young Serpens South star-forming region. 
Expanding on this idea, \citet{Pavlik2020} argues that observations of young star-forming regions suggest star clusters are born fully mass-segregated. 
More recently, ALMA observations by \citet{Xu2024} confirmed primordial mass segregation in eight massive, luminous, and blue-profile Galactic protoclusters.

Complementarily, \citet{tarricq2022}, using {\it Gaia} EDR3 data of $389$ open clusters, found that older clusters tend to be more mass-segregated than younger ones, likely due to a combination of segregation and evaporation effects.
Recent studies have added understanding to the complex relation between cluster age and mass distribution. 
\citet{pang2024} conducted an analysis of the present-day Mass Function, (PDMF) for 93 star clusters using Gaia DR3 data. 
They found that the power-law index remains stable at $\alpha \approx 2$ for clusters younger than $200$ Myr, consistent with the \citet{kroupa2001} IMF, but decreases significantly for older clusters, particularly when considering stars within the half-mass radius. 
This behavior of the power-law index with cluster age provides evidence for dynamical evolution effects, where low-mass stars are preferentially lost through processes such as two-body relaxation and tidal stripping.

In this work, we use Gaia-based high-quality estimated stellar masses and member stars of 15 selected nearby open clusters, to empirically build their mass-luminosity relation down below 0.5 $M_\odot$. 
With our mass estimates for all member stars per cluster, we examine both their mass segregation and PDMF. 
Our investigation is presented in this document, which is organized into six sections, summarized as follows. Section \ref{sec:data} presents the Gaia DR3 \texttt{Astrophysical parameters} catalog, from which we retrieve estimated stellar masses for as many member stars as possible.
It also details the selection of 15 open clusters with diverse ages and metallicities included in this study. 
Section \ref{sec:methods} comprises four subsections: (i) identification and separation of binary stars from main sequence single stars, (ii) a brief overview of how cubic spline fitting was used to extrapolate stellar masses for single stars not listed in the Gaia DR3 \texttt{Astrophysical parameters} catalog, (iii) estimation of mass for each component of the binary stars, and (iv) the application of the minimum spanning tree method to address the mass segregation problem.
Section \ref{sec:results} presents the outcomes for each cluster regarding mass segregation and the PDMF. 
Section \ref{sec:discussion} explores the advantages and limitations of the methodology employed, along with the implications and significance of our findings. 
Finally, we summarize our key conclusions and insights in Section \ref{sec:conclusions}.

%------------
\section{Data}
\label{sec:data}

With the exception of IMF studies such as \citet{Kirkpatrick2024} within 20 pc of the Sun, the low-mass stellar regime remains poorly sampled, hindering a robust characterization of the IMF in star clusters.
The {\it Gaia} DR3 catalog, with completeness down to $G \sim 21$ mag, includes a significant number of sub-solar stellar masses.
However, their astrometric and photometric errors are significant enough to impact and bias estimates of key physical parameters, such as distance, stellar mass, or the spatial distribution of a cluster.
The catalog provided by \citet{alfonso2024} contains candidate member stars for 370 nearby open clusters using Gaia DR3 data, considering stars as faint as $G \sim 19.5$ mag (see Fig. 5 in \citealt{alfonso2024}).
Their robust membership extends to sufficiently faint stars, providing a high-quality, complete sample of low-mass coeval stars, making it suitable for studying their IMF. 
From the open cluster sample presented in \citet{alfonso2024}, which includes stars with masses reported in the Gaia DR3 \texttt{Astrophysical parameters} catalog, we select those clusters with at least 100 stars with estimated masses. 
This absolute threshold is adopted to ensure sufficient statistical robustness for the analysis of mass segregation and the present-day stellar mass function.
To avoid including clusters with significant elongation along the line of sight due to parallax biases \citep{bailer2015,luri2018}, we exclude them through a visual inspection of their spatial distribution. 
Our final sample consists of 15 open clusters, their metallicity, age and distance modulus are listed in Table \ref{tab:parameters}. 
Several nearby clusters exhibit rather round cores and no discernible parallax-error stretch, but among the farthest clusters, the line-of-sight stretching was about 3 times the size of the distribution in the direction perpendicular to it.

\begin{figure*}[h]
\centering
\includegraphics[width=1.0\textwidth]{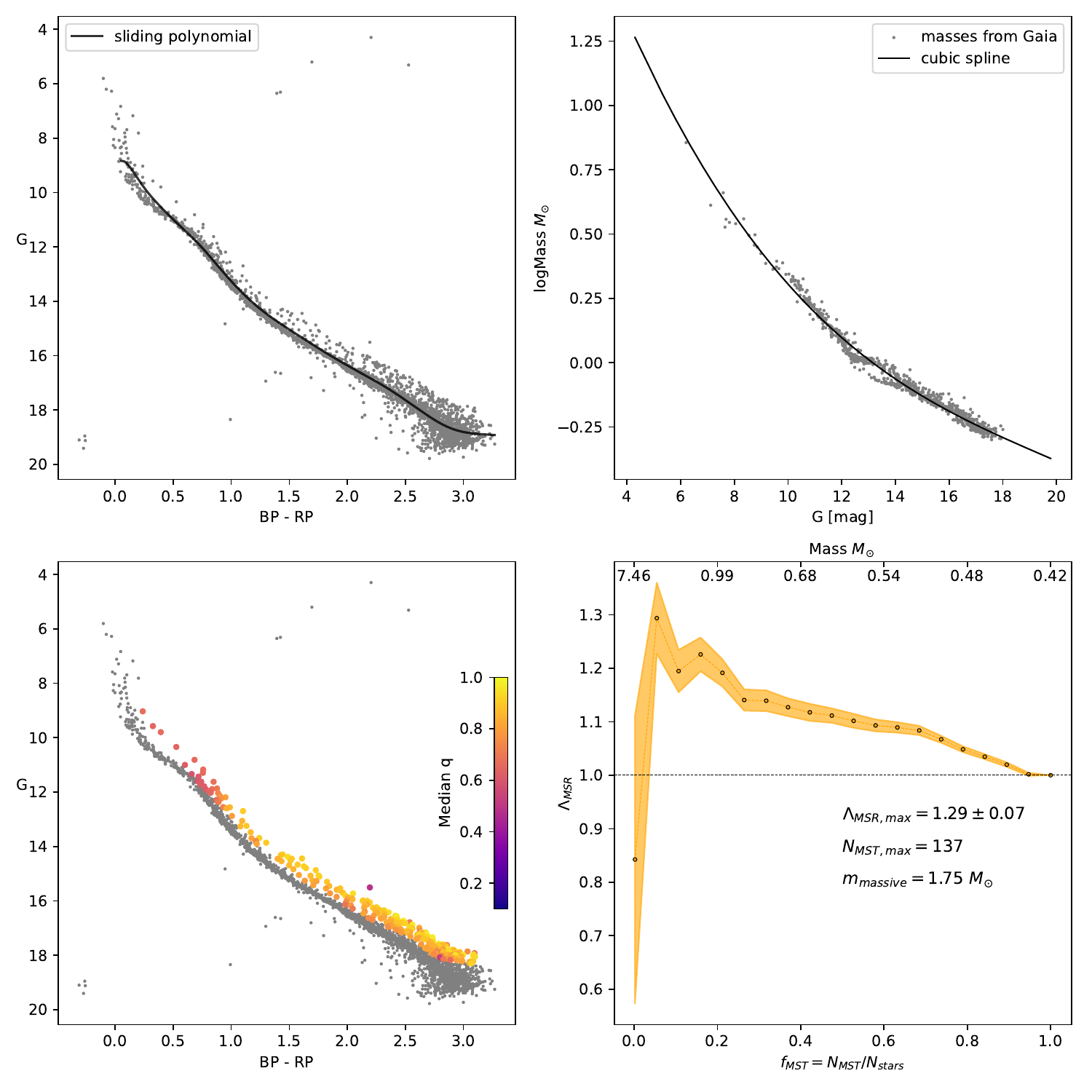}
\caption{Analysis of the open cluster NGC~2516 using the members provided in \citet{alfonso2024}.  
From left to right and top to bottom:  
CMD with the sliding polynomial used to select binary stars; spline fit to the stellar masses from the Gaia DR3 \texttt{Astrophysical Parameters} catalog; CMD with the mass ratio $q$ derived through the SBI method; and the minimum spanning tree ratio $(\Lambda_{\mathrm{MSR})}$ as a function of the fraction of stars used in the MST $(f_{\mathrm{MST}})$.}
\label{fig:full_trumpler_10}
\end{figure*}

The Gaia DR3 \texttt{Astrophysical Parameters} catalog is an additional resource to {\it Gaia} that provides key information such as temperature, radius, mass, age, luminosity, stellar surface gravity, and some chemical abundances.  
To estimate the mass of each star, the BaSTI stellar models \citep{hidalgo2018} are combined with atmospheric parameters ($T_{\text{eff}}$, log $g$, [$M/H$]) for stars with $G<18.25$. 
The mass estimates are generated using the Final Luminosity Age Mass Estimator (FLAME), which applies a Monte Carlo bootstrap method to sample each parameter. 
The estimated mass is taken as the $50_{\text{th}}$ percentile of the distribution, while the lower and upper uncertainties correspond to the $16_{\text{th}}$ and $84_{\text{th}}$ percentiles, respectively (see Section 11.3.6 in \citet{gaiadr32022} for further details). 
A cross-match using {\it Gaia} \texttt{source\_id} with Gaia DR3 \texttt{Astrophysical parameters} catalog found that a total of $43,386$ stars have a measured \texttt{mass\_flame} parameter. 
From the cross-match, we look for stars in the main sequence using \texttt{evolstage\_flame} 
$\leq 360$.
We note that the \texttt{mass\_flame} reported error sizes are almost always below $10\%$ of the estimated (median) mass. 
The minimum mass possible estimated is $0.5 M_\odot$.

\renewcommand{\arraystretch}{1.3}
\begin{table*}
\centering
\begin{tabular}{@{\extracolsep{1.3pt}}ccccccccccccc@{}}\hline\hline
\multirow{2}{*}{Cluster} & \multirow{2}{*}{$N_{\text{stars}}$} & \multirow{2}{*}{$N_{\text{binaries}}$} & \multirow{2}{*}{$\alpha$} &
\multirow{2}{*}{$\alpha_{\text{Pang}}$} & $m_{\text{MS,total}}$ & $m_{\text{lower}}$ & $m_{\text{upper}}$ & Z & {{$DM$}} & Age & Age$_{\text{CG}}$ \\
\cline{6-8} \cline{9-9} \cline{10-10} \cline{11-12}
 & & & & & \multicolumn{3}{c}{\(M_\odot\)} & \multicolumn{1}{c}{(dex)} & (mag) & \multicolumn{2}{c}{Myr}\\ \hline
Pozzo 1 & 626 & 79 & $2.24^{+0.09}_{-0.09}$ & - & 472.9 & $0.48$ & $10.06$ & -0.0690 &  7.600 & 11.6 & 9.5 \\
NGC 2547 & 465 & 63 & $2.25^{+0.16}_{-0.16}$ & $1.80 \pm 0.12$ & 387.4 & $0.51$ & $5.13$ & -0.0130 & 7.989 & 27.6 & 32.4 \\
IC 2602 & 462 & 60 & $1.51^{+0.13}_{-0.13}$ & $1.79 \pm 0.15$ & 325.4 & $0.37$ & $4.39$ & -0.0153 & 5.870 & 33.1 & 36.3 \\
Trumpler 10 & 929 & 116 & $2.06^{+0.11}_{-0.11}$ & - & 767.4 & $0.47$ & $5.95$ & 0.0450 & 8.149 &  40.1 & 32.4 \\
BH 99 & 544 & 64 & $2.05^{+0.12}_{-0.12}$ & $2.28 \pm 0.14$ & 450.4 & $0.47$ & $5.62$ & 0.0672 & 8.350 &  45.5 & 95.5 \\
Alessi 5 & 340 & 38 & $2.04^{+0.15}_{-0.15}$ & $1.95 \pm 0.23$ & 294.1 & $0.46$ & $6.39$ & 0.0632 & 8.141 &   48.9 & 67.6 \\
NGC 7058 & 313 & 44 & $2.03^{+0.13}_{-0.14}$ & $1.84 \pm 0.27$ & 205.5 & $0.35$ & $3.39$ & 0.1153 & 7.731 &  78.8 & 40.7 \\
Blanco 1 & 579 & 68 & $2.39^{+0.10}_{-0.10}$ & $2.15 \pm 0.11$ & 336.0 & $0.27$ & $2.74$ & -0.0155 & 6.900 &  79.5 & 104.7 \\
Melotte 22 & 1130 & 98 & $2.09^{+0.11}_{-0.11}$ & $2.01 \pm 0.09$ & 723.3 & $0.36$ & $4.35$ & 0.0287 & 5.554 &  103.5 & 77.6\\
NGC 6475 & 1091 & 106 & $2.02^{+0.07}_{-0.07}$ & $1.53 \pm 0.09$ & 1009.3 & $0.46$ & $5.92$ & 0.0201 & 7.279 &  271.0 & 223.9 \\
ASCC 101 & 154 & 17 & $1.63^{+0.17}_{-0.16}$ & - & 131.7 & $0.37$ & $5.66$ & 0.0043 & 8.080 &  276.7 & 489.8 \\
Alessi 9 & 267 & 28 & $1.63^{+0.20}_{-0.17}$ & $1.41 \pm 0.31$ & 199.1 & $0.40$ & $4.49$ & -0.0013 & 6.659 &  302.4 & 281.8 \\
NGC 3532 & 2206 & 249 & $1.89^{+0.07}_{-0.07}$ & $1.71 \pm 0.06$ & 2052.0 & $0.45$ & $4.82$ & 0.0584 & 8.479 &  651.6 & 398.1 \\
NGC 2516 & 2520 & 261 & $2.04^{+0.05}_{-0.05}$ & $2.02 \pm 0.07$ & 1998.9 & $0.44$ & $8.87$ & 0.0047 & 8.127 &  674.8 & 239.9 \\
NGC 2632 & 862 & 78 & $2.36^{+0.12}_{-0.11}$ & - & 565.6 & $0.41$ & $2.67$ & 0.1978 & 6.309 & 946.3 & 676.1 
\\
\hline
\end{tabular}
\caption{PDMF parameters for fifteen open clusters.
The number of member stars in each open cluster, $N_{\mathrm{stars}}$, is taken from \citet{alfonso2024}.
$N_{\mathrm{binaries}}$ denotes the number of binary systems identified using the sliding polynomial method with $q \geq 0.5$. The power-law index of the PDMF derived in this work and that reported by \citet{pang2024} are denoted by $\alpha$ and $\alpha_{\mathrm{Pang}}$, respectively. The total mass of main-sequence stars in each cluster, $m_{\mathrm{MS,\,total}}$, is
estimated from the spline-based masses, accounting for the individual components of binary systems. The lower and upper mass limits of the main-sequence stars used to compute the power-law fit to the PDMF are denoted by $m_{\mathrm{lower}}$ and $m_{\mathrm{upper}}$, respectively, and are indicated by vertical gray dashed lines in Figures. ~\ref{fig:imf_all_clusters} and ~\ref{fig:cdf}. The metallicity $Z$ and the distance modulus without correction by extinction {$DM$} are adopted from \citet{alfonso2024}. The quantities listed as Age and Age$_{\mathrm{CG}}$ correspond to the values reported in \citet{alfonso2024} and by \citet{cantat-gaudin2020}, respectively.
}
\label{tab:parameters}
\end{table*}

\section{Methods}
\label{sec:methods}

In this section, we describe the methodology used on the collected stellar masses data to extend such estimates to all stars in the selected clusters, considering the special case of binary systems.

%---------

\subsection{Identifying the binary stars with the sliding polynomial}
\label{sec:binaries_id}

Binary stars are present in the fifteen open clusters studied and can be identified in the observed $G$ vs. $BP-RP$ color–magnitude diagram (CMD) as a sequence of stars located above the main sequence, where $G$ corresponds to the apparent magnitude and $BP-RP$ to the reddened color as measured by Gaia \citep{malofeeva2022}.
In unresolved binaries, the combined flux of both components produces a magnitude offset relative to single stars of the same color that depends on the mass ratio, reaching $\sim 0.75$ mag for equal-mass systems \citep{tauris2023,cordoni2023}. 
This dependence produces the characteristic broadening and brightening of the main sequence.

To identify the main sequence single stars and separate them from the binary ones, we used a sliding polynomial function as proposed by \citet{stock1988}. 
In this fitting procedure, a moving polynomial is fitted around any given point by considering only the data within a fixed interval, and the data closer to the point is given a larger weight. 
In our case, we fitted a straight line to the CMD values, with the color data weighted by

\begin{equation}\label{equation:weights}
    w_i = \left( 1 - \left(\frac{|C_i - C|}{dC}\right)^2 \right)^{n/2}
\end{equation}
with $dC$ being the window size around color C, $n$ an integer, and $C_i$ the $i$-th star's color in the window.
We use $n=3$ and $dC=0.5$ mag. 
Fig. \ref{fig:full_trumpler_10} upper left panel shows the fit obtained for cluster NGC 2516.
The larger number of single stars and their concentration along the main sequence ties the sliding weighted straight line fit to it. 
The sliding polynomial provides a non-parametric description of the single stars' main-sequence, which is clearly concentrated in the CMD, whereas binaries follow a broader and flatter distribution, sometimes appearing as a distinct brighter sequence. 
Since the number of stars per color bin is not uniform—clusters naturally contain more faint stars than bright ones—the method applies a moving window that adjusts locally to the density of points. 
Within each window, the sliding polynomial fits the main sequence trend while accounting for the local spread in G-band magnitude. 
The visual inspection of all the clusters CMD showed that the dispersion of the single star's main sequence at each color is well characterized by $1.7$ times the median absolute deviation of the residuals within its window, typically about $0.2$ magnitudes. 
We also imposed the condition $G > 8.5$ magnitudes to avoid classifying subgiants and red giants as binary stars. 
Fig. \ref{fig:full_trumpler_10} lower left panel shows identified binary stars with colored symbols for cluster NGC 2516.

%---------

\subsection{Mass estimates for single stars using splines}
\label{sec:splines}

The collected \texttt{mass\_flame} values from the Gaia DR3 \texttt{Astrophysical parameters} catalog were used to produce mass estimates on stars lacking this parameter. 
To do so, we fitted a spline function on the known masses using the \texttt{SciPy} python library \citep{scipy2020}.
A spline is a piece-wise polynomial function \citep{deboor1978}, commonly used for interpolating problems. 
Since it is defined using polynomials, the ones on the two extremes of the data points or knots can be formally extended beyond, therefore acting as an extrapolation, and probably close to the real function if it is not too far from the extreme knots. 
We used the spline function \texttt{splrep} in \texttt{SciPy} with a cubic polynomial. 
First, we fitted the spline on the logMass versus G magnitude relation with a smoothing condition of $s=0.1$.
This condition controls the amount of smoothness in the sum of squared residuals, smaller values yield smoother fits.
Then, the \texttt{BSpline} function in \texttt{SciPy} was used to perform extrapolation based on the tuple containing the vector of knots and the spline coefficients from \texttt{splrep}.

Fig. \ref{fig:full_trumpler_10} upper right panel shows the spline fit (black line) obtained for NGC 2516 to estimate the mass for each of the $929$ stars members found in \citet{alfonso2024} based on the $254$ ones with \texttt{mass\_flame} values (gray points). 
These masses from {\it Gaia} reach down to $0.5$ \(M_\odot\), while in the extrapolation process, they are extended down to about $0.3$ \(M_\odot\) in the best cases.

Five of the fifteen open clusters have a few subgiants (SG) and red giants branch (RGB) stars. 
Alessi 9 has one SG or RGB with no \texttt{mass\_flame}, 
NGC 2516 has five RGBs of which one has \texttt{mass\_flame},
NGC 2632 has seven SGs of which six have \texttt{mass\_flame}, 
NGC 3532 has six RGBs of which three have \texttt{mass\_flame}, and
NGC 6475 has one SG or RGB with no \texttt{mass\_flame}.
In the case of NGC 2632, their spline-based masses are very close to the {\it Gaia} \texttt{mass\_flame} ones, but in other clusters some significant differences were found, with the spline-based mass being larger. 
In all these SG and RGB cases, the spline-based mass was used on the mass segregation calculations, since their order by mass is not altered significantly. 
On the other hand, for the  PDMF slope, these stars are omitted to avoid the bias their overestimated spline-based masses may introduce at the extreme right side of the PDMF. In this investigation, our focus is to extend the PDMF towards low mass values.

\begin{figure*}[t]
\centering
\includegraphics[width=1.0\linewidth]{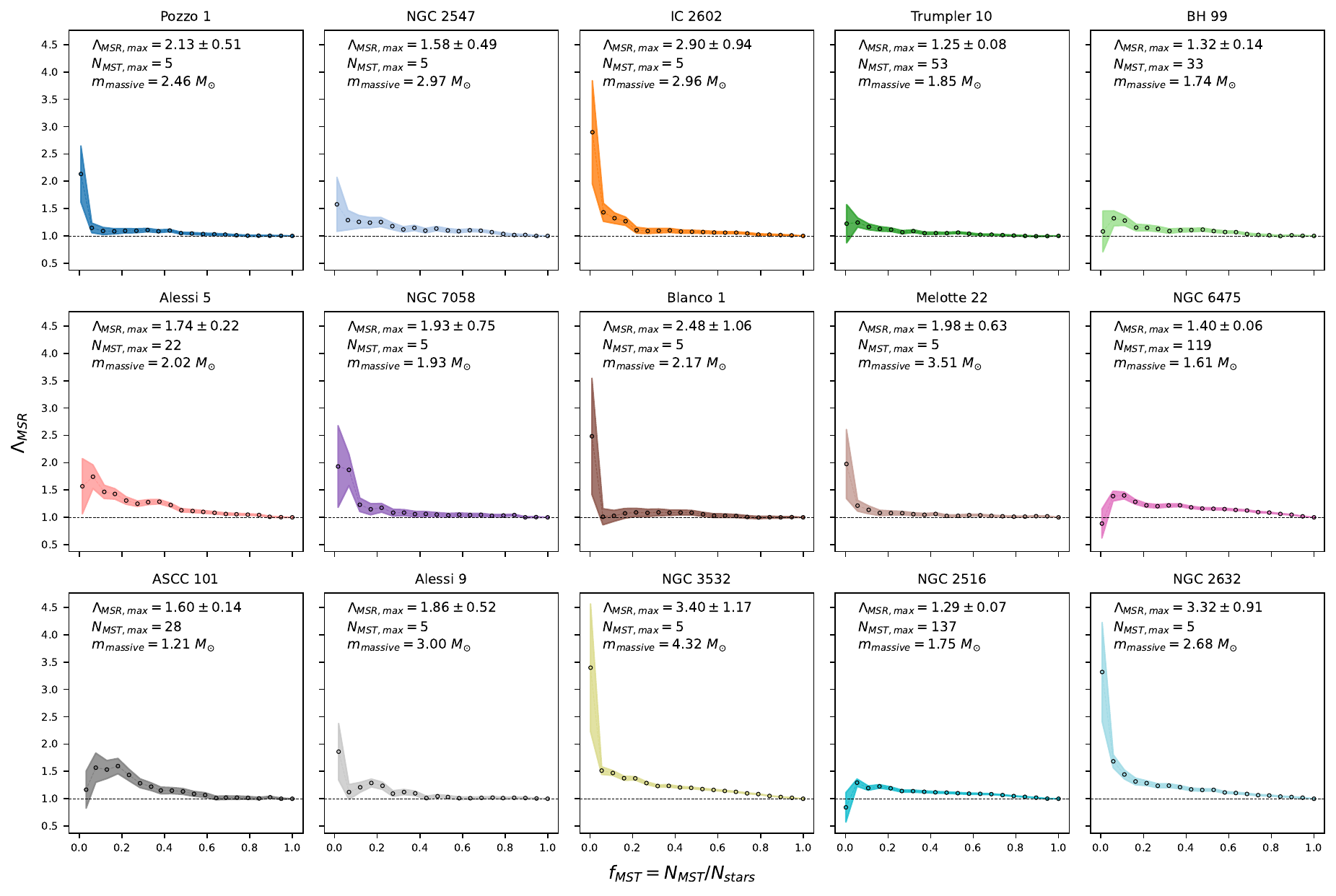}
\caption{
Mass segregation results. $\Lambda_{\mathrm{MSR}}$ as a function of the fraction of stars used in the MST ($f_{\mathrm{MST}}$) for the fifteen open clusters studied. 
The horizontal line at $\Lambda_{\mathrm{MSR}} = 1.0$ marks the regime with no mass segregation. The shaded regions represent $\sigma_{\mathrm{random}}$ computed at each $N_{\mathrm{MST}}$.
The legend of each subplot indicates:
$\Lambda_{\mathrm{MSR,max}}$; the value of $N_{\mathrm{MST}}$ at which the mass segregation ratio reaches its maximum;
$N_{\mathrm{MST,max}}$, which identifies the number of most massive stars for which the strongest mass segregation is found;
and $m_{\mathrm{massive}}$, the stellar mass threshold above which all stars are included in the computation.
}
\label{fig:mass_segregation_all_clusters}
\end{figure*}

%---------

\subsection{Individual masses of binary star components derived from Simulation-Based Inference}
\label{sec:binaries_masses}

Although masses for all single stars in each cluster are estimated from the spline fit, for the identified binary stars we compute the most probable individual masses of each component. 
To this end, we employ a Simulation-Based Inference (SBI) approach
\citep{cranmer2020}, following the methodology proposed by \citet{wallace2024}.

The SBI framework is built upon a vector of simulated magnitudes $\vec{x} = \{G, G_{BP}, G_{RP}\}$, generated with the \texttt{isochrones} package \citep{morton2015} using the MIST stellar models \citep{dotter2016,choi2016} in the \textit{Gaia} passbands.
In this work, all binaries are treated as hard systems, whose physical properties are described by the parameter vector $\vec{\theta} = (m_1, q, \tau, Z, d)$, where $m_1$ is the mass of the primary component, $q = m_2/m_1$ is the mass ratio, $\tau$ is the age, $Z$ is the metallicity, and $d$ represents the distance to the star.

The inference relies on prior assumptions for the astrophysical parameters $\vec{\theta}$. Following \citet{wallace2024}, we adopt beta distributions as priors for the primary mass $m_1$ and the metallicity $Z$, while uniform priors are assumed for the mass ratio $q$, the age $\tau$, and the distance $d$. 
Given these priors and the simulated photometry, the SBI method produces an approximation to the posterior distribution of $\vec{\theta}$ for each star, from which the most probable parameter values and their associated uncertainties are inferred. 
An example of the resulting posterior distributions is shown in the corner plot of Fig.~4 in \citet{wallace2024}.

The effect of unresolved binarity on the observed photometry depends strongly on the mass ratio $q$. As $q$ increases, the system departs progressively from the single-star main sequence, becoming redder and slightly brighter. 
The secondary component is also assumed to be a main-sequence star, with luminosity proportional to its mass; when it is less massive than the primary, it contributes a redder flux. 
In the limiting case $q = 1$, the binary consists of two identical stars and appears approximately $0.75$ magnitudes brighter than a single star of the same mass (see Fig.~1 in \citealp{malofeeva2022}).

In practice, the SBI method operates by generating mock photometric data from a set of input parameters and training a neural network to learn the mapping between the parameter space and the observables. 
To recover the astrophysical parameters of binary stars, we follow the same procedure as \citet{wallace2024}, simulating a training set of $10^{5}$ stellar systems using the \texttt{isochrones} package \citep{morton2015} to feed the model.
Once trained, the model is applied to the \textit{Gaia} photometry of binary candidates in each cluster to infer their posterior distributions. 
For each star, we compute the most probable values of the inferred parameters.

As an illustrative example, the lower-left panel of Fig.~\ref{fig:full_trumpler_10} shows the mass ratios inferred with the SBI method for the open cluster NGC~2516, while Fig.~\ref{fig:binary_estimations_sbi} displays the inferred mass ratios for all clusters. 
As expected, larger values of $q$ are found for stars exhibiting larger displacements from the main sequence.

Taking into account the photometric uncertainties reported by \textit{Gaia}, we adopt the criterion proposed by \citet{wallace2024}, whereby a system is classified as a binary if $q \geq 0.5$. Below this threshold, binary systems are unlikely to be reliably resolved through photometry alone, as their brightness enhancement above the main sequence lies within the observed photometric dispersion. 
For stars with inferred mass ratios below this limit, the spline-based mass estimate is therefore retained. 
When the system satisfies the binary criterion, the secondary mass is computed as $m_2 = q\,m_1$. 
The method adopted in this work is primarily sensitive to low mass-ratio systems located above the main sequence in the low-mass regime. 
This behavior may be driven by increased astrometric uncertainties toward the faint end of the \textit{Gaia} catalog \citep{rybizki2022,alfonso2024}. 
In addition, \citet{wallace2024} showed that the inferred parameters are sensitive to the adopted priors on age and metallicity. 
A fully self-consistent treatment would require drawing the primary and secondary masses according to a canonical initial mass function \citep[e.g.,][]{kroupa2001}, as discussed by \citet{Dabrin2022}. 
A detailed exploration of these effects is beyond the scope of the present work. 
Nevertheless, despite the reduced sensitivity of the method at the faintest
magnitudes, we retain systems satisfying $q \geq 0.5$ as binary candidates.

%-----------

\subsection{Mass segregation using the minimum spanning tree}
\label{sec:minspantree}

The Minimum Spanning Tree (MST) is a graph that connects a set of points with the minimum total edge length and no closed loops \citep{spanningbook2004}. 
In this work, the points correspond to the three-dimensional positions (X, Y, Z) of the stars in each cluster, and the MST is constructed in this spatial coordinate space. 
Stellar positions are derived from Gaia astrometry ($\alpha$, $\delta$ and distance $d=1 / \varpi$) and transformed into Galactic Cartesian coordinates using the \texttt{Astropy} package \citep{astropy2013,astropy2018,astropy2022}.
The MST has been widely used to detect the mass segregation effect in star clusters by comparing the total length of the MST formed by the most massive stars with that of MSTs constructed from randomly selected stars \citep{angelo2025,hu2025}.
In order to quantifying the mass segregation effect, we follow the methodology described by \citet{allison2009}. 
This method constructs the MST of the $N$ most massive stars ($N_{MST}$) and compares it to the MST of several samples of $N_{MST}$ random stars.
For a cluster with mass segregation, it is expected that the length of the $N$ most massive stars is shorter than the average length of the $N$ random stars. 
On the other hand, if the length of the $N$ most massive stars is greater than the average length of $N$ random stars, the cluster exhibits inverse mass segregation. 
A cluster with non-segregation thus contains a similar length of the MST for the massive stars and the random sample.

The quantification of the mass segregation using the MST proceeds as follows:

\begin{itemize}
    \item[$\bullet$] Sorts the data in descending order according to the mass of the stars. Then, compute the length of the $N_{MST}$ most massive stars ($l_{\text{massive}}$).
    \item[$\bullet$] Compute the average length and dispersion of a set of $N_{MST}$ random stars $\left\langle l_{\text{random}}\right\rangle \pm \sigma_{\text{random}}$ (assuming a Gaussian dispersion). 
    \item[$\bullet$] Determine the mass segregation ratio ($\Lambda_{MSR}$) and its uncertainty as
    \begin{equation}
        \Lambda_{MSR} = \frac{\left\langle l_{\text{random}}\right\rangle}{l_{\text{massive}}} \pm \frac{\sigma_{\text{random}}}{l_{\text{massive}}}
    \end{equation}
    \item[$\bullet$] Repeat the previous steps for different values of $N_{MST}$. Finally, plot $\Lambda_{MSR}$ versus $N_{MST}$ to observe the level of mass segregation in the cluster.
\end{itemize}

The number of random realizations used to sample reference stars is not arbitrary.
Numerical experiments have shown that using $50$ or more random sets provides reliable estimates of the associated uncertainties \citep{allison2009}. 
In this work, we find that adopting $300$ random realizations yields stable results. The minimum spanning tree (MST) is computed in Galactic Cartesian coordinates $(X, Y, Z)$ using the \texttt{SciPy} Python package \citep{scipy2020}, specifically the \texttt{minimum\_spanning\_tree} routine based on Kruskal’s algorithm \citep{kruskal1956}. 
Mass segregation is quantified through the minimum spanning tree ratio, $\Lambda_{\mathrm{MSR}}$, which is evaluated as a function of the fraction of cluster members included in the tree. Starting from the most massive stars, progressively less massive members are incorporated until all identified cluster members are included.
The analysis is performed at exactly twenty discrete values of the fraction of stars included in the MST. 
Each point shown in the figures corresponds to a cumulative subset of cluster members ordered by decreasing stellar mass.  
The first point represents the case in which only the five most massive stars are included, while the twentieth (final) point corresponds to the inclusion of all cluster members, such that the number of stars in the MST equals the total number of members reported by \citet{alfonso2024}. 
This procedure is illustrated for the NGC~2516 cluster in the lower-right panel of Fig.~\ref{fig:full_trumpler_10} and applied to the full cluster sample shown in Fig.~\ref{fig:mass_segregation_all_clusters}. 
For the purpose of computing mass segregation, binary systems are treated as single objects with a total mass equal to $m_1 + m_2$.

%-----------
\section{Results}
\label{sec:results}

%-----------
\subsection{Mass Segregation}
\label{sec:mass_segregation}

An analysis of the $\Lambda_{\mathrm{MSR}}$ versus $f_{\mathrm{MST}}$ profiles presented in Fig.~\ref{fig:mass_segregation_all_clusters} reveals two distinct trends, in which no age tendencies were found. 
The first group are clusters that show a smooth increase in $\Lambda_{\mathrm{MSR}}$ reaching a maximum within the interval $f_{\mathrm{MST}} \in (0, 0.2)$, followed by a gradual and continuous decline as $f_{\mathrm{MST}}$ increases, without exhibiting abrupt variations. 
This behavior is observed in Trumpler 10, BH 99, Alessi 5, NGC 6475, ASCC 101, and NGC 2516. 
These six clusters show only mild signatures of mass segregation, suggesting that the spatial distribution of stars of different masses remains relatively uniform throughout the cluster volume.

The second behavior is observed in the Pozzo 1, IC 2602, Blanco 1, and Melotte 22 clusters, which show a sudden drop from $\Lambda_{\mathrm{MSR}} \gtrsim 2$ to values close to unity within the interval $f_{MST} \in (0,0.2)$ where the derivative changes its sign. 
After that, the profiles stabilize on a plateau around $\Lambda_{\mathrm{MSR}} \sim 1$ from $f_{MST} > 0.4$.
Clusters NGC 3532 and NGC 2632 exhibit the same sudden drop but, unlike the others, they reach a plateau at a slower rate. 
It is noticed that the NGC 7058 and Alessi 9 clusters have a different behavior, in which the drop is not as deep as the previous ones. 
In all cases, this slump indicates a strong concentration of massive stars, specifically the five most massive, in the core regions of these eight clusters, consistent with significant mass segregation, as illustrated in Fig.~\ref{fig:mass_segregation_all_clusters}. 
Finally, the cluster NGC 2547 exhibits a mixed behavior between smooth and sudden drops, having an intermediate mass segregation level.

\begin{figure}
\centering
\includegraphics[width=1.0\linewidth]{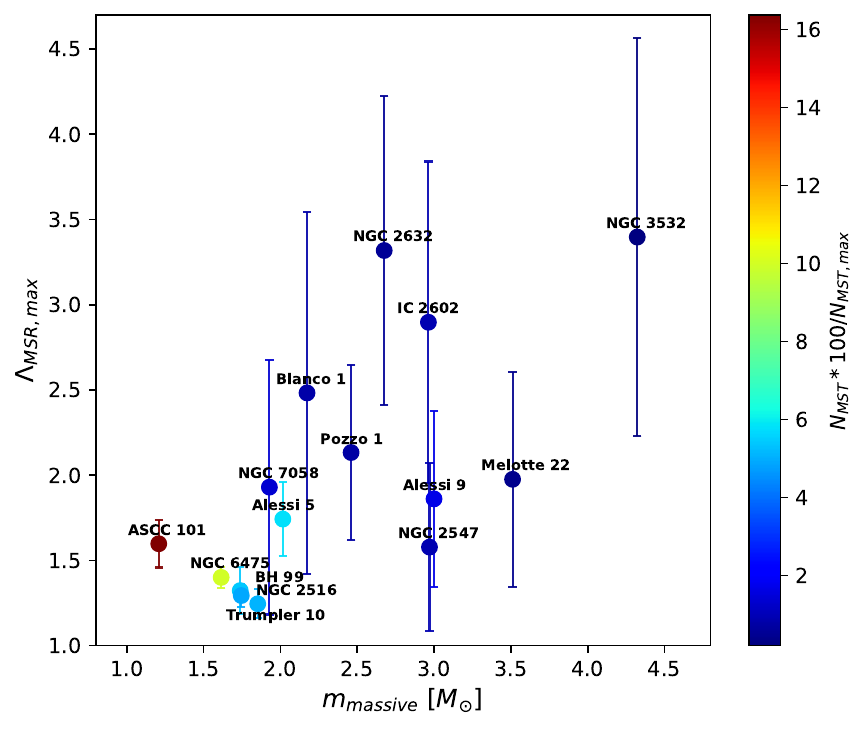}
\caption{$\Lambda_{\mathrm{MSR,max}}$ versus $m_{\mathrm{massive}}$ computed from Fig. \ref{fig:mass_segregation_all_clusters}, 
showing the corresponding error bars. The color bar indicates the percentage of $N_{\mathrm{MST}}$ 
for which the maximum value of $\Lambda_{\mathrm{MSR,max}}$ is reached, normalized by the number of stars in the cluster reported in Table~\ref{tab:parameters}. 
The larger error bars correspond to cases with $N_{\mathrm{MST}} = 5$, 
where the MST algorithm naturally yields higher uncertainties.}
\label{fig:lambda_max}
\end{figure}

An analysis of the values of $N_{MST}, \Lambda_{MSR,max}$ and $m_{massive}$ labeled each panel in Fig. \ref{fig:mass_segregation_all_clusters}, reveals a trend which is shown in Fig. \ref{fig:lambda_max}. 
When the number or fraction of segregated stars is large ($N_{MST}>20, N_{MST}/N_{stars}>5\%$), the amount of segregation is low ($\Lambda_{MSR,max}<1.8$) and $m_{massive}\lesssim 2 M_\odot$. 
When the number or fraction of segregated stars is low ($N_{MST}=5, N_{MST}/N_{stars}<2\%$), the amount of segregation $\Lambda_{MSR,max}>1.5$ and $m_{massive}\gtrsim 2 M_\odot$. 
Cluster NGC 3532 has the lowest percentage of segregated stars (0.23\%), the highest amount of segregation ($\Lambda_{MSR,max}=3.40 \pm 1.17$) among the most massive stars ($m_{massive}=4.32 M_\odot$) of all clusters studied. 
This translates into a slight correlation between the amount of segregation and the mass above which it occurs, with a Pearson correlation coefficient of 0.67 between $m_{massive}$ and $\Lambda_{MSR,max}$. 
In the sample studied, 7 clusters have $m_{massive}\lesssim 2 M_\odot$ and 
$\Lambda_{MSR,max}<2.0$, 5 have $m_{massive}>2 M_\odot$ and $\Lambda_{MSR,max}>2.0$, and only 3 have $m_{massive}>2 M_\odot$ but $\Lambda_{MSR,max}<2.0$ .

These numbers suggest that in clusters with not-so-massive stars, mass segregation is less intense and acts over a larger number of stars; and in clusters with very massive (and also very few) stars, mass segregation is limited to only these few stars, and it is frequently (but not always) strong.
This is an expected trend, considering that the larger the stellar masses, the larger the gravity they exert.

In our sample, we find both young and old clusters exhibiting strong segregation among their five most massive stars ($m_{\mathrm{massive}} > 2\,M_\odot$): Pozzo 1, IC 2602, and Blanco 1 (young) versus NGC 3532 and NGC 2632 (old).  
Conversely, there are also young and old clusters showing only weak segregation, involving values greater than $5\%$ of the stars with $m_{\mathrm{massive}} < 2\,M_\odot$: Trumpler 10 and BH 99 (young) versus NGC 2516 (old).  
This suggests that the degree of segregation is more strongly driven by the mass of the most segregated stars than by the cluster age.

Also, in the young and old clusters with intense segregation group above, a difference is observed in how $\Lambda_{MSR,max}$ drops afterwards: in younger clusters, this number drops and is flatter at around 1, while in older ones, it drops and decreases slowly towards 1. 
In the younger clusters, this intense segregation points towards a scenario of primordial segregation, while for the older ones, we may argue that dynamical mass segregation acting over time may have reached lower mass stars but at a lower intensity.

Finally, there is the consideration of mass loss for the most massive stars, which can happen very early in the life of a cluster. 
The two oldest clusters of this study have white dwarfs. 
The oldest one, NGC 2632, has the most, 12, for which our estimated masses are not only incorrect but, more importantly, in the past, they were the most massive stars in the cluster. 
Estimating their current mass is beyond the scope of this paper. Nonetheless, we visually inspected their current distribution in each of the XYZ coordinates and compared those to the whole cluster and to the 35 most massive stars subset. 
NGC 2632's whit dwarfs look like the cluster, therefore we can speculate that after becoming white dwarfs very early in the history of NGC 2632, dynamical mass segregation has acted on them with their current lower mass, pushing them away as time passes, then reversing their original concentration when they were massive stars. 
Measuring their initial mass segregation status is impossible because two-body relaxation has - by definition - erased the previous dynamics and therefore positions.

%----------

\subsection{The Present-Day Mass Function}
\label{sec:imf}

To characterize the Present-Day Mass Function (PDMF) of each cluster, we adopt a power-law function in its simplest form, $\Phi(m)\propto m^{-\alpha}$, as used by \citet{pang2024}. 
$\Phi(m)$ corresponds to the normalized-by-area histogram of the number of stars per mass bin and $\alpha = 2.35$ is referred to as the \citet{salpeter1955}'s slope. 
In the appendix Section \ref{appendix:pareto}, we explain how $\Phi(m)$ is mathematically equivalent to the probability density function (PDF) of a truncated Pareto distribution. The associated PDF to the PDMF is then

\begin{equation}\label{equ:pdf}
\text{PDF }(m \mid \alpha) =
\frac{ (\alpha-1) m_{\text{lower}}^{(\alpha-1)} m^{-\alpha} }
{ 1-\left(\frac{m_{\text{lower}}}{m_{\text{upper}}}\right)^{(\alpha-1)} }
\end{equation}

To derive a likelihood function that can be used to make an inference on $\alpha$, we define, given a sample of stellar masses $\{ m_i \}_{i=1}^N$, the log-likelihood as

\begin{equation}
\log \mathcal{L}(\alpha) = \sum_{i=1}^N \log \text{PDF }(m_i \mid \alpha).
\end{equation}

By substituting the expression for the PDF into the log-likelihood, we obtain the explicit function

\begin{align}\label{equ:log_likelihood}
\log \mathcal{L}(\alpha) = \,
& N \log (\alpha-1) + N (\alpha-1) \log m_{\text{lower}} \notag \\
& - \alpha \sum_{i=1}^N \log m_i \notag \\
& - N \log\left( 1 - \left( \frac{m_{\text{lower}}}{m_{\text{upper}}} \right)^{(\alpha-1)} \right).
\end{align}

\begin{figure*}%[h]
\centering
\includegraphics[width=1.0\linewidth]{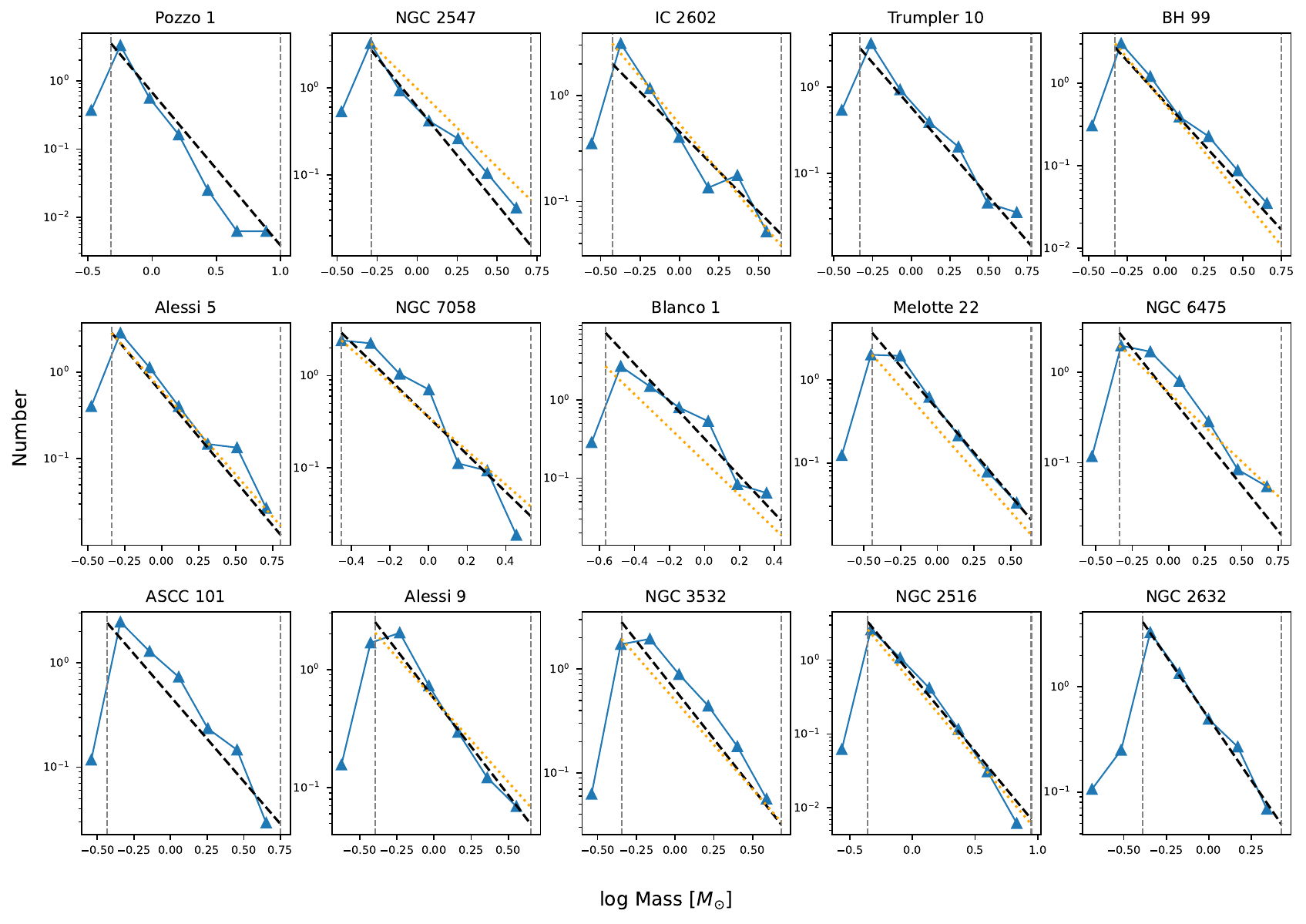}
\caption{
The PDMF of the fifteen open clusters studied in this work. Blue triangles and solid lines show the PDMFs derived from the masses of all main-sequence stars, including both single and binary systems. For a direct comparison with
\citet{pang2024}, we adopt the same binning scheme used in their analysis.
The black dashed line in each panel represents the power-law fit described in
Sect.~\ref{sec:imf}, computed within the lower and upper stellar mass limits listed in
Table~\ref{tab:parameters}, which are indicated by the vertical dashed gray lines.
The orange dotted lines show the power-law fits reported by \citet{pang2024}.
The corresponding power-law indices $\alpha$ derived in this work and those reported by
\citet{pang2024} are listed in Table~\ref{tab:parameters}, except for the clusters
Pozzo~1, Trumpler~10, ASCC~101, and NGC~2632.
}
\label{fig:imf_all_clusters}
\end{figure*}

The function in equation~\eqref{equ:log_likelihood} allows us to estimate the parameter $\alpha$ that best characterizes the PDMF, by maximizing the agreement between the model and the observed mass distribution, restricted to main sequence stars only.
In this work, we use the stellar masses derived from the spline interpolation and the binaries treatment, in other words, for the stars classified as binaries, each component's mass is taken separately. 
We also adopt $m_{\text{lower}}$ as the location of the low-mass inflection point in the Cumulative Distribution Function (CDF), and $m_{\text{upper}}$ as the maximum observed stellar mass for main sequence stars. 
These values are listed in Table~\ref{tab:parameters} and are represented as vertical dashed lines in Figs.~\ref{fig:imf_all_clusters} and ~\ref{fig:cdf}.

Then, to look for the $\alpha$ that maximize the log-likelihood function \eqref{equ:log_likelihood}, we minimize the negative of the log-likelihood function performing the \texttt{Nelder-Mead} algorithm using the \texttt{minimize} method in \texttt{SciPy} \citep{scipy2020}.
Second, to estimate uncertainties on these parameters, we implement an MCMC sampling using the \texttt{emcee} Python package \citep{foreman2013}. 
The sampling was performed with one hundred random walkers and two thousand iterations.
A uniform prior on the $\alpha$ parameter was used, and the lower and upper uncertainties were chosen as the $16_{\text{th}}$ and $84_{\text{th}}$ percentiles. 
We validate the convergence of the chains by looking at the \citet{gelman1992} diagnostic $\hat{R}$, which compares the variance within chains to the variance between chains.
We confirm that for all clusters, values of $\hat{R} \leq 1.2$ were obtained, indicating that the convergences have been reached.

Table \ref{tab:parameters} lists the PDMF power-law parameters, along with their associated uncertainties, for the fifteen open clusters in our sample. 
Figure \ref{fig:imf_all_clusters} shows the corresponding PDMFs and the fitted power-law functions. 
The mean PDMF slope across the sample is $-2.14 \pm 0.26$. 
No correlation is found between metallicity and the PDMF power-law index, likely due to the limited metallicity range spanned by the sample.
Also, we did not see any significant difference in the mass distribution slope with age. 
For all the younger clusters with main sequence stars only (from Pozzo 1 to Melotte 22, age $\lesssim$ 100 Myr), the obtained mass distribution is probably very close to their IMF.

%---------
\section{Discussion}
\label{sec:discussion}

%-------
\subsection{Mass segregation}

We first compare and discuss our results with previous studies for three individual clusters: Blanco~1, Melotte~22, and NGC~2516. 
The open cluster Blanco~1 was previously studied by \citet{zhang2020}, who reported evidence of mass segregation among cluster members more massive than $1.4\,M_\odot$, with a mass segregation ratio of $\Lambda_{\mathrm{MSR}} \sim 1.5$. 
This result is consistent, within the uncertainties, with the value $\Lambda_{\mathrm{MSR}} \sim 2.5$ obtained in this work for the nine most massive stars.

For Melotte~22 (the Pleiades), we find significant mass segregation for the $20$ most massive stars, with $\Lambda_{\mathrm{MSR}} \sim 2.0$, in agreement with the results reported by \citet{raboud1998} and \citet{converse2008}. 
In addition, \citet{heyl2022} identified a skewed distribution of low-mass stars escaping beyond the cluster tidal radius, a feature commonly associated with mass segregation.
Although \citet{heyl2022} did not explicitly compute the minimum spanning tree, the presence of mass segregation is evident from the fact that longer escape times correspond to lower-mass stars, as shown in Fig.~9 of their study. 
In the case of NGC~2516, \citet{pera2022} reported a mass segregation ratio of $\Lambda_{\mathrm{MSR}} \sim 1.4$ for the ten most massive stars, in good agreement with our results. 
By comparing the cluster age with its relaxation time, they concluded that the observed segregation is primordial rather than dynamical in origin. 
However, the age adopted by \citet{pera2022} ($\sim100$~Myr) is significantly younger than the ages reported by \citet{alfonso2024} and \citet{cantat-gaudin2020}, which may affect the interpretation of the segregation timescale.

On a larger scale study, \citet{dib2018} obtained a skewed distribution of mean $\Lambda_{MSR}$ considering only $N_{MST}\leq 10$ for $1276$ Galactic clusters, in which $180$ clusters were found segregated with $\Lambda_{MSR} > 1.5$ indicating mass segregation in all of them. 
For 13 out of our 15 clusters (missing only BH 99 and Pozzo 1), \citet{dib2018} found the above mentioned mean value of $\Lambda_{MSR}$ to be larger than $ 1.0$, which is consistent with our results.

The typical timescale for dynamical mass segregation driven by two-body interactions is the relaxation time $t_{\text{relax}}$, which for a stellar cluster depends on the number of member stars and its crossing time $t_{\text{cross}}$, as follows from \citet{binney2008}

\begin{equation}
t_{\text{relax}}=\frac{N_{\text{stars}}}{8\ln{N_{\text{stars}}}}t_{\text{cross}} \, ,
\end{equation}
with $t_{\text{cross}}$ estimated from the diameter divided by a typical velocity dispersion taken as $1$ km/s, or equivalently $1.023$ pc/Myr. 
\citet{alfonso2023} measured the tidal radius of Blanco 1, Melotte 22, and NGC 2632, as $13.97$, $11.35$, and $12.12$ pc, respectively. 
Considering the number of stars $N_{\text{stars}}$ found for these clusters in \citet{alfonso2024}, and taking twice the tidal radius as the diameter of the cluster, the estimated relaxation times of these clusters are $311$, $446$, and $378$ Myr, respectively (rounding to the closest integer). 
As a result, Blanco 1 and Melotte 22 have lived only $0.25$ of their relaxation time but NGC 2632 has already lived $2.5$ times its $t_{\text{relax}}$. 
These clusters showed no trend on mass segregation vs $t_{\text{relax}}$. 
The fact that both Blanco 1 and Melotte 22 already show mass segregation at such early time would support the scenario of a primordial mass segregation for them.

In a cluster that is old enough, dynamical mass segregation will occur regardless of the initial degree of segregation. 
The longer dynamical mass segregation operates, the more effective it becomes at concentrating the most massive stars, with the effect acting more rapidly on higher-mass members \citep{zwart2010}. 
Our results support this conclusion. For the sample studied, the mass of the most massive stars appears to be the dominant factor controlling the strength of mass segregation, while increasing dynamical times allow the segregation signature to extend toward progressively lower-mass stars, albeit with reduced intensity.

A final comment on our results is that the selected sample of fifteen clusters out of 370 available from \citet{alfonso2024}, was chosen because they looked not too stretched in 3D space, due to {\it Gaia} parallax uncertainties. 
Theoretically, the MST approach to mass segregation could overcome this effect and still be able to measure segregation if present, despite the obvious deformation in the shape of the cluster. 
Here, we have tested that MST works in the more nominal-looking ones. Overall, our findings do not contradict what is qualitatively expected for mass segregation.

%--------
\subsection{Binary disruption by two-body collisions}

The same two-body collisions that cause mass segregation among single stars cause binary star disruption \citep{vesperini2011,bromley2012}. 
Therefore, the effects of two-body collisions over time on binary stars are also worthy of consideration.
Three widely recognized mechanisms are responsible for breaking binary systems: (i) supernova explosion of one of the stars, (ii) close interactions with other stars, and (iii) unbinding wide binaries \citep{binney2008,ciotti2021}.

What outcome arises first, mass segregation or binary disruption, is hard to ascertain, but we see that the percentage of binary stars is lower for the older than 100 Myr clusters, while younger ones have generally higher percentages (see Fig. \ref{fig:percentage_binary_stars}).
The small difference between the minimum (9\%) and maximum (14\%) binary percentage in the clusters studied, and the small size of the sample are limitations to reach a firm conclusion on this matter. 
More clusters need to be studied to give statistical robustness to this result.
However, this points to the fact that older clusters would have fewer binary stars, because of their being disrupted over time. 
We did not see any correlation between the fraction of binaries and the values of $N_{MST,max}, \Lambda_{MSR,max}$ or $m_{massive}$.

%----------
\subsection{Present Day Mass Function}

The PDMF shape of the star clusters shown in Fig. \ref{fig:imf_all_clusters} could support the theoretical predictions made by \citet{Haghi2015}. 
These authors investigated 100 million years of evolution in modeled clusters and demonstrated that the ejection of residual gas from initially mass-segregated clusters leads to a significantly shallower PDMF slope. 
The observed break in the low-mass regime of the PDMF for most of the clusters shown in Fig. \ref{fig:imf_all_clusters} may suggest evidence of an early ejection of gas from initially mass-segregated clusters. 
In contrast, the small change in the PDMF slope observed in NGC 7058 could be consistent with the results of \citet[see Fig. 1]{Haghi2015} for clusters that lack primordial mass segregation. 
Together, these findings would suggest that a large fraction of the open clusters in the Milky Way could have primordial mass segregation.

Given the limited amount of data and the errors associated to the lower mass fainter stars, an alternative explanation can simply be the incompleteness of the photometric data. 
There is also an error associated with the spline-based masses just above the above-mentioned break, because of photometric errors. 
In Appendix \ref{appendix:mass_vs_g}, we explore this issue, using cluster NGC 3532 as an example. We estimate that at $m_{\text{lower}}$, photometric errors induce a 5-10\% error in the spline-based masses.

In relation to the PDMF $\alpha$ values, we perform a direct comparison between our results reported in Table~\ref{tab:parameters} and those presented by \citet{pang2024} in their Table 1. 
Eleven clusters are common to both samples; for these, \citet{pang2024} report a mean $\alpha = 1.86 \pm 0.15$, while our analysis yields $\alpha = 1.99 \pm 0.24$ for the same clusters. 
Both values are consistent within the uncertainties and fall within the error bars of our overall result for the full sample of fifteen open clusters, $\alpha = 2.01 \pm 0.25$. 
This latter value is also consistent, at the $1\sigma$ level, with the canonical power-law index of $2.3\pm 0.36$ reported by \citet{kroupa2001}.

A recent review on the physical origin of the IMF by \citet{hennebelle2024} summarizes how different mechanisms can yield primordial variations of the IMF, particularly below 0.5 M$_\odot$, where the slope could reduce to zero and even become positive below 0.2 M$_\odot$ (see their Fig. 3). 
 Regarding the possibility that our clusters' PDMF may not be truly characterized by a single power law fit but rather a piecewise function, particularly at the low mass regime, our results in Fig. \ref{fig:imf_all_clusters} show some slight variations that could be or not significant, given the size of our sample. 
 In principle, if not biased by incompleteness, PDMF at the low-mass regime should be their IMF since these stars have not lost any mass yet. 
 But, if mass segregation has been strong enough or has acted long enough to eject low-mass members, including disrupted binary stars' secondary masses, the PDMF slope at the low-mass regime will be flatter than at higher mass. 
 Separating the effect of sustained mass segregation from primordial variations of the IMF slope is complicated by the uncertain age estimations against which to compare their relaxation time.

%-----------
\subsection{Binaries components mass estimation}

The isochrones employed to assess the masses of the stars by the SBI treatment \citep{cranmer2020} are not the same as the ones used by {\it Gaia} \citep[BaSTI]{hidalgo2018}. 
This is not ideal but it is a good enough first approach. 
In any case, there is still room for improvement, given that robust estimates of distance, age, and metallicity are available for our cluster members from \citet{alfonso2024}, consequently, we only need inferences on $m_1$ and $q$. 
We are currently developing an improved procedure that incorporates priors on $m_1$ and $q$ for future analyses.

A second consideration is what coordinates $X$, $Y$ and $Z$ are assigned to the secondary stars in the binary systems, as their location is relevant for the mass segregation effect. 
After the binary correction using the SBI approach in Sect. \ref{sec:binaries_masses}, the new estimated masses require Galactic cartesian coordinates to be included in the process of computing the MST. 
Conversely, identifying the possible positions of the two components requires additional information, such as the system's period, and in our case, we only have their masses. 
However, the separation of binary stars does not exceed $3000$ AU \citep{Jerabkova2019} even in the case of wide binaries, which are rare in open clusters \citep{deacon2020}. 
For the closest cluster in our study, Melotte 22, this difference translates into a maximum parallax deviation for the secondary of less than one microarcsecond with the primary's parallax, clearly below the parallaxes uncertainties in the {\it Gaia} data. 
In any case, assigning the same position to both binary components, as we do, does not affect the calculation of the MST for the mass segregation effect.

In order to evaluate how sensitive are the PDMF slope values to the binaries treatment, we considered two extreme cases for the individual masses of the binary star components. 
First, we considered the binary stars as single and used their spline-based mass, and second, we entirely omitted the binary stars.
In the first case, the PDMF slope values were within 10\% of our results for 13 clusters.
In the second case, results were more biased towards lower values of $\alpha$ down to 15\% except for 2 clusters.

By taking a binary star as a single one, we count two lower-mass stars as one of higher mass. 
In other words, we are missing two stars of lower mass and adding one of higher mass. 
On the other option, i.e., ignoring the binary star from the mass distribution altogether, we are missing two stars of lower mass.
In both cases, the effect is to make the PDMF slope flatter, i.e. $\alpha$ is reduced. 
Since our identified binaries account for only 10\%$-$15\% of all the $N_{\text{stars}}$ members of the cluster, ignoring the binaries reduces the total number of stars by at most twice this factor, mainly at lower mass values. 
The first procedure looks less harmful. 
If such biased and forceful treatments did not visibly alter the results, then our current, more sophisticated approach must not be far from true, uncertainties permitting.

Another potential source of systematic uncertainty arises from the presence of rapidly rotating stars in open clusters. 
Such objects have been identified spectroscopically in open clusters in both the Milky Way and the Magellanic Clouds. 
A fraction of these rapidly rotating stars can exhibit the Be phenomenon and typically span spectral types from late O through B to early A \citep{Rivinius2013}. 
In this context, \cite{Brandt2015,BD2015} used rotating stellar models to construct synthetic color–magnitude diagrams, showing that cluster sequences can be reproduced assuming a single stellar population with a range of rotation rates and viewing angles. Observationally, rotation has also been linked to broadened and, in some cases, split main sequences in young and intermediate-age clusters \citep{Cordoni2024}.  
Based on stellar models, \cite{Girardi2019} found that the deviations of fast rotators relative to non-rotating stars are small, but non-negligible. 
Owing to our magnitude cut at $G>8.5$ mag, early-type stars, where the effects of rapid rotation are most pronounced, constitute only a small fraction of our sample. 
Therefore, the stellar mass estimates are not expected to be significantly affected by the presence of fast rotators. 
This is consistent with Fig. \ref{fig:imf_all_clusters}, which shows that the number of stars with masses above $2\,M_\odot$ is very small. 
Although stellar rotation can modify the morphology of the upper main sequence, these effects are not expected to introduce significant biases in our stellar mass estimates or in the derived mass-ratio distribution.

\begin{figure}
    \centering
    \includegraphics[width=1.0\linewidth]{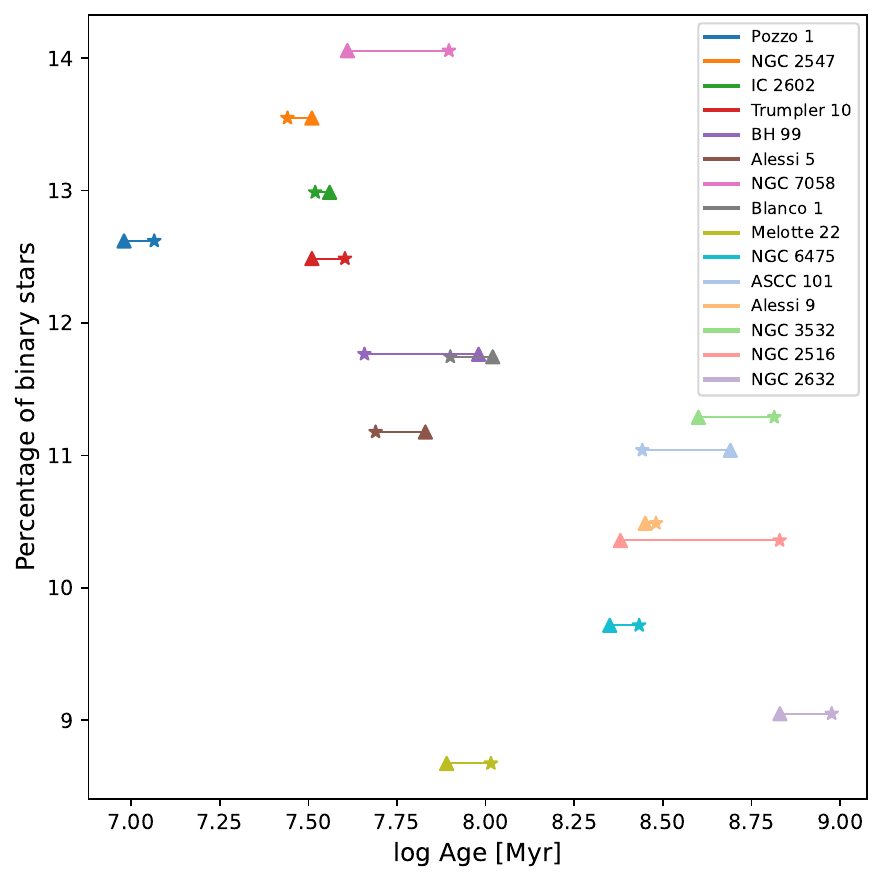}
    \caption{Percentage of binary stars per cluster ($N_{\text{binaries}}/N_{\text{stars}} \times 100\%$) versus the $\log(\text{Age})$ values from \citet{alfonso2024} (stars) and \citet{cantat-gaudin2020} (triangles) for the fifteen open clusters studied. The line segments indicate the differences between the $\log(\text{Age})$ values reported for each cluster.}
    \label{fig:percentage_binary_stars}
\end{figure}

%---------
\subsection{Revision of cluster ages from previous literature}

All the previous discussion is based on the ages of the open clusters as estimated by \citet{alfonso2024}. 
In our previous investigation, we noticed a trend between our ages log(age)$_{\text{Alfonso}}$ and those of \citet{cantat-gaudin2020} log(age)$_{\text{CG}}$, for older clusters (see Fig. 9 upper panel in \citealt{alfonso2024}). 
More specifically, we noted that $\Delta$log(age) $=$ log(age)$_{\text{Alfonso}}$ - log(age)$_{\text{CG}}$ vs log(age)$_{\text{Alfonso}}$ becomes positive and reaches up to 0.5, for log(age)$_{\text{Alfonso}}\gtrsim 8.2$, i.e., age $\gtrsim$ 160 Myr.

For this reason, we also explored the trend we observed in the percentages of binary stars with age, as seen in Fig. \ref{fig:percentage_binary_stars}, using the ages log(age)$_{\text{CG}}$, listed also in Table \ref{tab:parameters}.
Our conclusion remains the same, the proportion of binary stars in the clusters diminishes with time, the only exception being Melotte 22. 
A larger sample will clarify is this cluster is a single deviant point or corresponds to a real trend in the population.

Another relevant point concerns the relation between the upper stellar mass ($m_{\mathrm{upper}}$) listed in Table \ref{tab:parameters} and cluster age. 
Since older clusters are expected to have lower main-sequence turn-off masses, a decreasing trend of $m_{\mathrm{upper}}$ with age is anticipated. 
Most clusters follow this behavior. 
An exception is NGC~2516, which shows a relatively high $m_{\mathrm{upper}}$ for the age estimated in \citet{alfonso2024}. 
However, a significantly younger age for this cluster was reported by \citet{cantat-gaudin2020}, which is more consistent with the $m_{\mathrm{upper}}$ value inferred from our main-sequence spline-based mass estimates.
We checked that for this cluster, the Gaia DR3 \texttt{Astrophysical parameters} catalog has a main sequence bright star, \texttt{source\_id}=5289522090708710656, with \texttt{mass\_flame}=7.1672807 Msun, that also has parameters calculated by the Extended Stellar Parametrizer for Hot Stars (ESP-HS) routine, which lists a spectral type \texttt{spectraltype\_esphs}=B for it. 
Keep in mind that $m_{\text{upper}}$ marks the most massive main-sequence star, and it contains this star in the calculation because of its value of \texttt{evolvstage\_flame}=267.

This apparent drawback gives support to the spline-based masses employed in this investigation and points us to a potential procedure to revise the open cluster ages from \citet{alfonso2024} and other references, using the Gaia DR3 \texttt{Astrophysical parameters} catalog and the mass distributions. 
In this way, the highest main sequence star spline-based mass based on \texttt{mass\_flame} must correlate with the estimated age, as expected from well-established results of stellar evolution. This insight will be considered for future investigations.

%-------------------
\section{Conclusions}
\label{sec:conclusions}

In this work, we have conducted a homogeneous analysis of mass segregation and the present-day stellar mass function in fifteen nearby open clusters using high-precision astrometric and photometric data from the \textit{Gaia DR3} mission. 
Mass segregation is detected, to varying degrees, in all the clusters studied. 
Two distinct segregation patterns are identified: one characterized by strong segregation confined to a small number of the most massive stars, and another involving weaker but more spatially extended segregation affecting a larger fraction of the stellar population. 
Our results indicate that the degree and spatial extent of mass segregation are primarily associated with the maximum stellar mass present in a cluster, while no clear dependence on cluster age is found within the analyzed sample. 
Strong mass segregation is observed in several clusters, including some at early evolutionary stages, suggesting that non-dynamical processes may contribute to the establishment of mass segregation, while older clusters also show signatures consistent with the progressive action of dynamical segregation toward lower-mass stars.

The present-day stellar mass function of the analyzed clusters is well described by a power law consistent at the one-sigma level with the canonical Kroupa initial mass function reported for massive stars. 
In most clusters, a break is observed in the low-mass regime of the mass function. Photometric incompleteness is likely a primary contributor to this feature; however, intrinsic variations in the low-mass shape of the stellar mass function, as well as the early dynamical evolution of initially mass-segregated systems, including the effects of gas expulsion or their combined action, may also play a role. 
In contrast, the mild slope change observed in NGC~7058 appears consistent with a non-mass-segregated origin. 
Additionally, older clusters show evidence of binary disruption, reflecting the cumulative impact of dynamical evolution over time. 
Together with the evidence for binary disruption in older clusters, these results highlight how both initial conditions and subsequent dynamical evolution jointly shape the stellar mass distribution of open clusters in the Milky Way.

%---------
\begin{acknowledgments}
The authors would like to express their sincere gratitude to the anonymous referee and to professors Pavel Kroupa, Hosein Haghi, Václav Pavlík, and Luis Aguilar for their valuable comments and feedback, which significantly improved this article. 
We also thank Artem Lutsenko for drawing our attention to the work that developed the methodology applied here for the simulation-based inference of binary stars. 
The authors would like to thank the Vice Presidency of Research \& Creation’s Publication Fund at Universidad de los Andes for its financial support and also the Fondo de Investigaciones de la Facultad de Ciencias de la Universidad de los Andes, Colombia, through its Programa de Investigación código INV-2025-213-3418. This work has made use of data from the European Space Agency (ESA) mission {\it Gaia} (\url{https://www.cosmos.esa.int/gaia}), processed by DPAC, (\url{https://www.cosmos.esa.int/web/gaia/dpac/consortium}). 
Funding for the DPAC has been provided by national institutions, in particular the institutions participating in the {\it Gaia} Multilateral Agreement.
\end{acknowledgments}

\newpage
\appendix 

\newpage
\section{E\lowercase{stimations of the $q$ ratio through} SBI}
%\section{Estimations of the q ratio through SBI}
\label{appendix:sbi_estimations}

\begin{figure*}[h]
    \centering
    \includegraphics[width=1.0\linewidth]{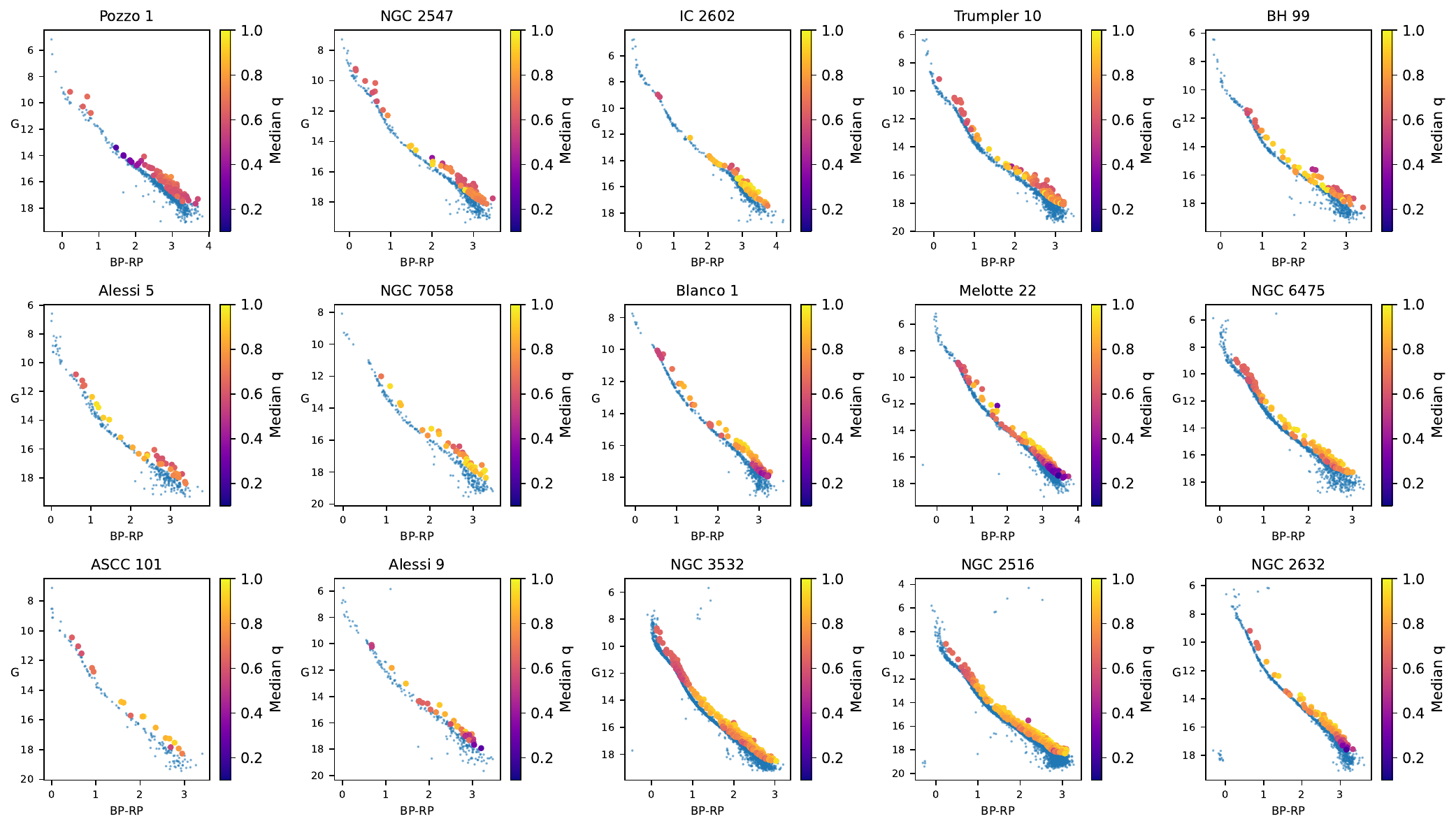}
    \caption{CMD showing the mass ratio $q$ estimated with the SBI method for binary star candidates selected from the sliding polynomial. Only stars with $q > 0.5$ are considered binaries; below this threshold, it is unlikely that a binary can be reliably identified based on photometry alone.}
    \label{fig:binary_estimations_sbi}
\end{figure*}

%---------
\newpage
\section{C\lowercase{umulative} D\lowercase{istribution} F\lowercase{unction}}
\label{appendix:cdf}

\begin{figure*}[h]
    \centering
    \includegraphics[width=1.0\linewidth]{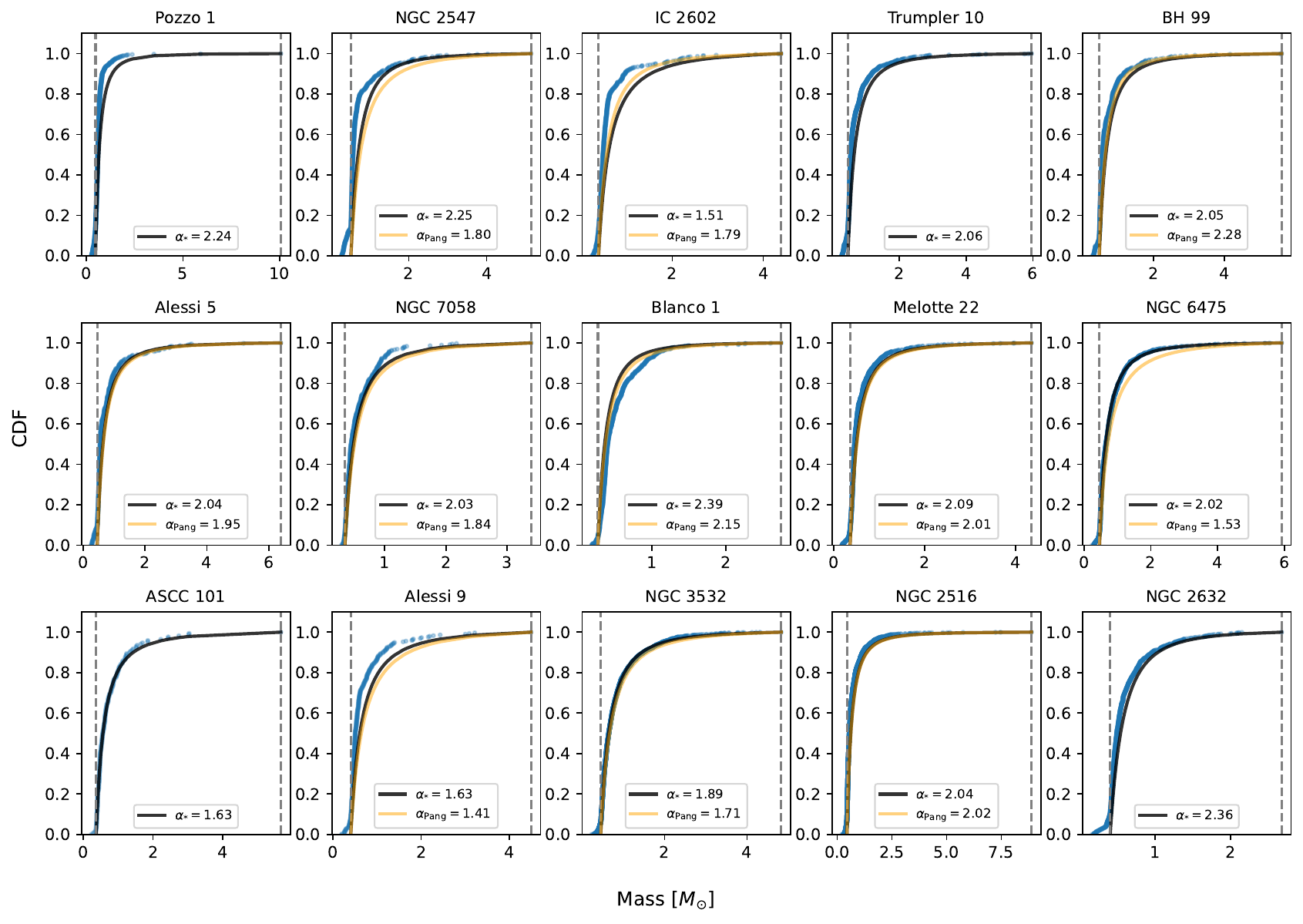}
    \caption{CDF for main-sequence stars only. The vertical dashed gray lines indicate $m_{\text{lower}}$ and $m_{\text{upper}}$ from Table~\ref{tab:parameters}, respectively. The black solid lines show the CDF with the $\alpha$ parameter estimated in this work, while the orange lines correspond to the $\alpha$ values reported by \citet{pang2024}, when available.}
    \label{fig:cdf}
\end{figure*}

%----------
\newpage
\section{P\lowercase{resent} D\lowercase{ay} M\lowercase{ass} F\lowercase{unction and the truncated} P\lowercase{areto distribution}}
\label{appendix:pareto}

If $N(m)$ is the cumulative distribution of the number of stars of mass $m$, then $N(m)$ describes how many stars up to mass $m$ are formed. 
Therefore, $\Phi(m)=dN/dm$ is the density distribution of the number of stars of mass $m$, which describes any histogram of the number of stars in the mass bin $(m,m+\Delta m)$. 
$\Phi(m)$ is generally described by a power law of the form $\Phi(m)\propto m^{-\alpha}$ and $\alpha=2.35$ is referred to as the \citet{salpeter1955}'s slope. 
When working with the histogram of $\log(m)$, as in Fig. \ref{fig:imf_all_clusters}, then
\[
\frac{dN}{dm}=\frac{dN}{d(\log(m))}\frac{d(\log(m))}{dm}=\frac{dN}{d(\log(m))}\frac{1}{\ln(10)m}
\Longrightarrow \frac{dN}{d(\log(m))} = \ln(10)m \frac{dN}{dm} \propto m \; m^{-\alpha} 
\Longrightarrow \frac{dN}{d(\log(m))} \propto  m^{-(\alpha-1)}
\]

%\[
%\Longrightarrow \frac{dN}{d(\log(m))} \propto  m^{-(\alpha-1)}
%\]

A normalized-by-area histogram of the number of stars per bin of mass, described by the PDMF, follows the power-law $\Phi(m)$. 
Since the PDMF is normalized in the interval $(m_{\text{lower}},m_{\text{upper}})$ then it is equivalent to the probability density function (PDF) of the truncated Pareto distribution in the interval $(m_{\text{lower}},m_{\text{upper}})$ with shape parameter $\alpha-1$. 
The shape of any histogram depends on the chosen bin size, but the data cumulative distribution is uniquely given by the data's percentile vs. the data, as seen in Fig. \ref{fig:cdf} (blue points). 
The cumulative distribution is particularly useful to obtain a precise determination of the inflection point of the PDMF at low mass, which sets the value of $m_{\text{lower}}$. 
As for $m_{\text{upper}}$, which corresponds to the highest-mass main sequence star per cluster, this was determined for each cluster by visual examination of the color-magnitude diagram. 
The PDF associated to the PDMF and its corresponding CDF are given by

\begin{equation}
\text{PDF }(m \mid \alpha) =
\frac{ (\alpha-1) m_{\text{lower}}^{(\alpha-1)} m^{-\alpha} }
{ 1-\left(\frac{m_{\text{lower}}}{m_{\text{upper}}}\right)^{(\alpha-1)} } \; \Longrightarrow \;
\text{CDF }(m \mid \alpha) = \int_{m_{\text{lower}}}^m \text{PDF }(m \mid \alpha)\; dm = 
\frac{1-\left(\frac{m_{\text{lower}}}{m}\right)^{(\alpha-1)}}
{1-\left(\frac{m_{\text{lower}}}{m_{\text{upper}}}\right)^{(\alpha-1)}}
\end{equation}

In general, the truncated Pareto distribution is defined for $a\leq x \leq b$ with shape parameter $c$ by

\begin{equation}
    \text{PDF }(x \mid a,b,c) = \frac{c a^cx^{-(c+1)}}{1-\left(\frac{a}{b}\right)^c}
    \hspace{1cm}\mbox{and}\hspace{1cm}
    \text{CDF }(x \mid a,b,c) = \frac{1-\left(\frac{a}{x}\right)^c}{1-\left(\frac{a}{b}\right)^c}\;\;\;.
\end{equation}
Its application to astrophysics, particularly to the masses of stars and the radii of asteroids, has been studied by e.g. \citet{zaninetti2008}. 
In this context $c=\alpha-1$.

%-------------

\newpage
\section{M\lowercase{ass estimations for the} NGC 3532 \lowercase{open cluster}}
\label{appendix:mass_vs_g}

\begin{figure*}[h]
\centering
\includegraphics[width=0.4\linewidth]{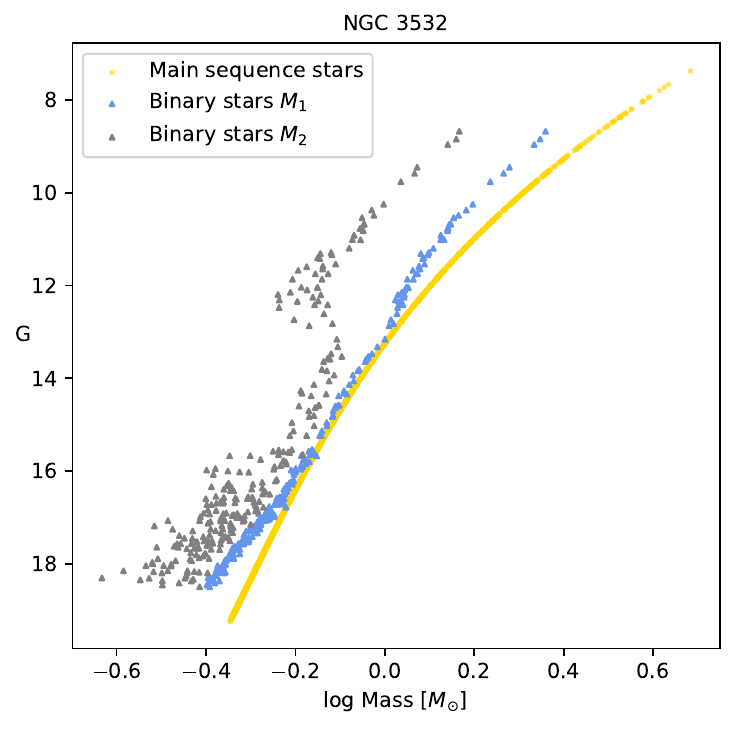}
\caption{
Mass estimates as a function of the observed \textit{Gaia} $G$ magnitude for main-sequence stars in the open cluster NGC~3532. 
The $G$ magnitude corresponds to the unresolved photometric measurement of each stellar system.
Masses of binary stars (blue triangles correspond to $M_1$ first component, while gray triangles correspond to $M_2$ second component) are derived using the Simulation-Based Inference (SBI) method described in Sect.~3, while masses of single stars (yellow stars) are obtained from the spline-based method.
For binary systems, the primary components define a tighter sequence that closely follows the single-star relation, whereas the secondary components exhibit a broader distribution toward lower masses. 
Both components of each binary are plotted using the same unresolved \textit{Gaia} $G$ magnitude.}
\label{fig:mass_vs_g}
\end{figure*}

In this section we briefly explain how the estimated masses of this work -- in particular, the SBI-based binary masses -- are distributed versus magnitude, for cluster NGC 3532. This is done as an example to illustrate what happens with all clusters, since it has a large number of binaries (249) spanning over 10 magnitudes. 
As seen in Fig. \ref{fig:mass_vs_g}, the single main sequence stars follow the yellow smooth curve, the binary stars' primary components form a tight sequence visible left of it (blue triangles), while the secondary ones follow a wider band to the left of the primaries (blue triangles). 
In this plot, both components of each binary are plotted with the same $G$ magnitude, their unresolved brightness from Gaia, so it is easier to identify them. 
The distance between the primary and secondary sequences gives information on the $q$-ratio computed by the SBI method and how it changes with magnitude $G$.

For the brightest binaries, the $q$-ratio is about 0.6 for $G\lesssim 12$ with some increased dispersion at $G\sim 12$. 
Then it steadily increases up to about 0.9 for $12\lesssim G \lesssim 15$, and for stars with $G\gtrsim 16$, the $q$-ratio spans from 0.6 to 0.9, showing a larger dispersion. 
Since color and mass are correlated with magnitude for main-sequence stars, then this behavior of the $q$-ratio is also replicated vs. color and the mass of the primary. 
In this cluster, most of the stars below $m_{\text{lower}}=0.45 M_\odot$ ($-0.35$ in log-scale) are secondary components (108 secondaries vs. 30 primaries). 
The way we identify the main sequence single stars (sliding polynomial), taking into account the effects of photometric errors (0.7 MAD of residuals), and how the spline smoothly continues the relation between $\log(Mass)$ and $G$ magnitude below $0.5 M_\odot$, translates into the continuation of the power-law for the stars identified as single. 
Binary stars in the faintest magnitude range will end up as two lower-mass stars, and the secondary ones will contribute the most below $m_{\text{lower}}$.
From this plot, it is also possible to estimate the effect of photometric errors on our spline-based mass of the single stars, at the faintest limit where we are extending the current Gaia's masses below $0.5 M_\odot$. 
The dispersion for the faintest and reddest part of the CMD is $\pm 0.5$ magnitudes. 
Given the slopes of the spline fits at $m_{\text{lower}}$, a $\pm 0.5$ magnitudes difference translates into a mass difference of about 
$5\%$. 
In NGC 3532, it means a mass difference of $0.025 M_\odot$. A more conservative photometric error estimation of up to 1 magnitude would translate into a mass difference of $10\%$.

%% For this sample we use BibTeX plus aasjournalv7.bst to generate the

\newpage
\bibliography{sample701}{}
\bibliographystyle{aasjournalv7}



\end{document}